\documentclass[aps,pre,showpacs,twocolumn]{revtex4-1}
\usepackage[utf8]{inputenc}
\usepackage[english]{babel}
\usepackage{graphicx}
\usepackage{amsmath}
\usepackage{amsfonts}
\usepackage{amssymb}

\begin{document}

\title{Universality and time-scale invariance for the shape of planar Lévy processes}
\date{February 7, 2014}
\author{Julien Randon-Furling}
\email{Julien.Randon-Furling@univ-paris1.fr}
\affiliation{SAMM (EA 4543), Universit\'e Paris-1 Panth\'eon-Sorbonne,\\ Centre Pierre Mend\`es-France, 90 rue de Tolbiac, 75013 Paris, France}

\begin{abstract}
For a broad class of planar Markov processes, viz. Lévy processes satisfying certain conditions (valid \textit{eg} in the case of Brownian motion and Lévy flights), we establish an exact, universal formula describing the shape of the convex hull of sample paths. We show indeed that the average number of edges joining paths' points separated by a time-lapse $\Delta \tau \in \left[\Delta \tau _1, \Delta \tau_2\right]$ is equal to $2\log \left(\Delta \tau_2 / \Delta \tau_1 \right)$, regardless of the specific distribution of the process's increments and regardless of its total duration~$T$. The formula also exhibits invariance when the time scale is multiplied by any constant.\\ Apart from its theoretical importance, our result provides new insights regarding the shape of two-dimensional objects (\textit{eg} polymer chains) modelled by the sample paths of stochastic processes generally more complex than Brownian motion. In particular for a total time (or parameter) duration~$T$, the average number of edges on the convex hull (``cut off'' to discard edges joining points separated by a time-lapse shorter than some $\Delta \tau < T$) will be given by $2 \log \left(\frac{T}{\Delta \tau}\right)$. Thus it will only grow logarithmically, rather than at some higher pace.
\end{abstract}

\maketitle

\section*{Introduction}
Over the last decades, many physical objects have been modelled by sample paths of stochastic processes. One of the best-known example remains perhaps polymer chains, with the Edwards, Rouse and Zimm models~\cite{Kac,Edw,dGe,West,DoiEdw}. Often, the outer shape of such physical objects plays a key rôle in their behaviour or function --- one may think for instance of the hydrodynamics of polymer fluids (see~\cite{FouDes} and references therein), or of the way the biological function of a protein will be linked to its shape~\cite{Cell,PRCDM}.

Mathematical elements allowing to describe the shape of stochastic processes' sample paths in two or three-dimensional space therefore have been, and still are, sought and needed. Probability distributions of quantities such as moments of inertia or asphericity~\cite{FouDes,Sciut,HRW} have been studied by physicists and biologists. In the mathematical literature, sustained attention, over the last fifty years, to the stochastic process associated with convex hulls of sample paths has led to numerous results, in particular concerning the average perimeter and area of these hulls~\cite{Lev,SW,Bax,Kin1,Ta,El,KBSM,CHM,Kho,Le2,SnS,Verz,Go,Go2,BiLe,KLM}. However, results pertaining to the number of facets or edges forming the boundary of the convex hull of a random sample remain rare when the sample does not consist of independent points~\cite{RS,RS2,Ef,Ald} or of the successive positions of a discrete-time random walk~\cite{Bax}. Only recently, the average number of edges on the convex hull of $n \geq 1$~independent, planar Brownian paths was computed~\cite{BMCHEdges}, but the derivation relied on specific aspects of the standard Brownian propagators and could not be extended to more general processes that have greater relevance for physical applications such as polymer chain modelling.

\begin{figure}
\begin{center}
\includegraphics[scale=0.3]{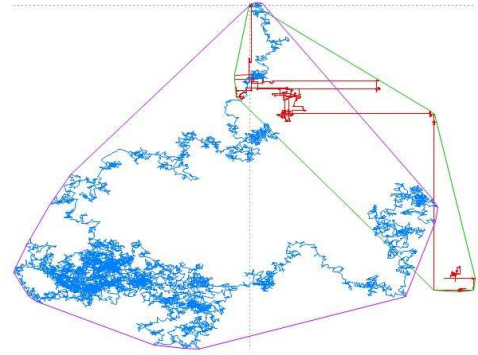}
\end{center}
\caption{\label{CHBMCP}Sample paths and their convex hulls for: a Cauchy-Lorentz process (right) and Brownian motion (left). Both paths have the same duration and were started from the origin (located at the top of the picture).}
\end{figure}

In this paper, we establish a general formula giving the average number of edges on the convex hull for a wide class of continuous-time stochastic processes in the plane. The derivation builds on elementary path transformations and on the uniform law of sojourn times in~$\mathbb{R}^{\pm}$ for certain one-dimensional processes tied to~$0$ at both ends~\cite{Kn,Kal,Yor,ABR}, which is a well-known result in the case of Brownian bridges.

A key point in our reasoning is to characterize edges by the time-lapse between the two points that they join on the process's path, as in~\cite{BMCHEdges}. Writing $\Delta \tau$ for this time-lapse, the formula we obtain is best expressed as giving the average number $\left\langle V \right\rangle$ of edges with $0<\Delta \tau _1 \leq \Delta \tau \leq \Delta \tau _2$:
\[ \left\langle V\left(\Delta \tau _1,\Delta \tau _2\right) \right\rangle = 2\log \left( \frac{\Delta \tau_2}{\Delta \tau_1} \right). \]
Remarkably, this formula is valid for Brownian motion, that has increments with finite mean, but also for heavy-tailed processes like Cauchy-Lorentz processes and other types of so-called Lévy flights, that have infinite-mean increments. Also, the formula exhibits some scale-invariance in time, and implies for example that, on average, for the convex hull to gain one edge, one has to prolong the duration of the process by about $60$~percent, if the same cut-off in time scale is used (NB: we are assuming that there is no limitation on the spatial scale at which we can see the process's path. Note incidentally that points separated by a small time-lapse are not necessarily close to one another in space, especially in the case of heavy-tailed processes.).
\medskip

We shall consider, generally, two-dimensional stochastic processes with stationary, independent increments. Such processes are called Lévy processes, and can be seen as a generalization in continuous time of a standard discrete-time random walk with independent, identically distributed jumps. We introduce our reasoning in the discrete-time case before extending it to continuous time and then comparing our analytical formula with the results of computer simulations.

\section{Discrete-time case}
\label{DT}
For the sake of clarity, let us first examine a discrete-time Markov process: a random walk in the plane with independent, identically distributed jumps. Its path will consist in $T=N$ steps ($N\geq2$). Let us denote its convex hull by~$\mathcal{C}_N$ and the boundary of~$\mathcal{C}_N$ by~$\partial \mathcal{C}_N$. An edge on $\partial \mathcal{C}_N$ will join points separated by $k$ steps, for some $1\leq k \leq N$; we call such an edge a $k$-edge. The question is: on average, how many $k$-edges will appear on the convex hull of the walk's path? We write $\langle V^{(k)} \rangle$ for this average number and we label $0$ through to $N$ the points on the random walk's path.

\subsection{Number of $k$-edges}
For $1\leq k \leq N$ and $0\leq i \leq N-k$, we call $\mathcal{L}^{i}_{i+k}$ the line joining the $i$-th and $(i+k)$-th points of the walk. Thanks to the Markov property, the three following events will be independent:
\begin{enumerate}
\item[(a)] the first part of the walk (from point $0$ to point $i$, i.e. the pre-$i$ part) does not cross $\mathcal{L}^{i}_{i+k}$, but it \textit{does} hit the line at step~$i$;
\item[(b)] between the $i$-th and the $(i+k)$-th step, the walk does not cross $\mathcal{L}^{i}_{i+k}$ but it \textit{does} visit the line at these steps;
\item[(c)] the third part of the walk (from point $i+k$ to point $N$, i.e. the post-$(i+k)$ part) does not cross $\mathcal{L}^{i}_{i+k}$, but it \textit{does} start from it at step~$i+k$.
\end{enumerate}
Labelling~$1$ and $2$ the sides of $\mathcal{L}^{i}_{i+k}$, one further requires, for each event, that the side on which the walk's steps remain is side~$1$.\\
Whence an expression of $\langle V^{(k)} \rangle$ in terms of: the probability $m^{0}_{(i)}$ associated with event (a), the probability $e^{(i)}_{k}$ associated with event (b), and the probability $m^{(i+k)}_{N}$ associated with event (c). Recalling that there are indeed two sides to a line, one obtains:
\begin{equation}
\langle V^{(k)} \rangle = 2\,\sum _{i=0}^{N-k} m^{0}_{(i)}\, e^{(i)}_{k}\, m^{(i+k)}_{N} \label{eq:Vk}
\end{equation}

If one assumes that the~$e^{(i)}_{k}$'s do not depend on~$i$ (which is the case for a walk with independent, identically-distributed jumps), then $e^{(i)}_{k}=e_k$ for all~$i$ and Eq.~(\ref{eq:Vk}) becomes:
 \begin{equation}
\langle V^{(k)} \rangle = 2\,e_{k}\, \sum _{i=0}^{N-k} m^{0}_{(i)}\, m^{(i+k)}_{N}. \label{eq:Vk2}
\end{equation}

The terms remaining under the summation sign are simple products of probabilities corresponding to \textit{independent} events: namely, the pre-$i$ part of the walk \textit{and} the post-$i+k$ part do not cross $\mathcal{L}^{i}_{i+k}$ other than at steps $i$ and $i+k$. Each term in the sum is therefore equal to the probability of the \textit{joint} event~$A^{i}_{k}$ (still conditioned on the fact that $\mathcal{L}^{i}_{i+k}$ is visited at steps $i$ and $i+k$), i.e.
\begin{equation}
m^{0}_{(i)}\, m^{(i+k)}_{N} = \mathbb{P}\left(A^{i}_{k}\vert B^{i}_{k}\right),\nonumber
\end{equation}
with $A^{i}_{k}=\left\lbrace \mathcal{L}^{i}_{i+k}\mathrm{\ not\ crossed\ before\ } i \mathrm{\ nor\ after\ }i+k \right\rbrace$ and $B^{i}_{k}=\lbrace \mathcal{L}^{i}_{i+k} \mathrm{\ is\ visited\ at\ steps\ }i \mathrm{\ and\ } i+k  \rbrace$.

Note that, thanks to the Markovian nature of the walk, with independent identically-distributed increments, the product $m^{0}_{(i)}\, m^{(i+k)}_{N}$ is the same as the probability that an identical walk with only $N-k$ steps stays on one side only of a line and hits it at time $i$ \textit{given} that it does hit the line at some point. The sum over $i$ going from $0$ to $N-k$ will therefore yield:
\begin{equation}
\sum _{i=0}^{N-k} m^{0}_{(i)}\, m^{(i+k)}_{N}=1,\nonumber
\end{equation}

Hence, inserting this into Eq.~(\ref{eq:Vk2}), one finds that:
 \begin{equation}
\langle V^{(k)} \rangle = 2\, e_{k}, \label{eq:Vk3}
\end{equation}
that is, the average number of $k$-edges appearing on $\partial \mathcal{C}_N$ is equal to twice the probability that a $k$-step random walk stays on a given side of the line joining its initial and final positions (this is called performing an excursion from the line) --~in particular, note that:
 \begin{equation}
\langle V^{(k)} \rangle \leq 2. \nonumber
\end{equation}
Note also that $V^{(k)}$ does not depend on $N$, the total number of steps in the walk: this can be understood if one realizes that as $N$ grows, the probability for a given $k$-segment to appear as an edge on the walk's convex hull becomes smaller, certainly, but at the same time there are more $k$-segments on the walk's path.

\subsection{Discrete edge-excursion formula}
Since $\langle V_N \rangle$, the average number of edges on $\partial \mathcal{C}_N$, can be expressed as:
\begin{equation}
\langle V_N \rangle = \sum_{k=1}^{N} \langle V^{(k)}\rangle,
\end{equation}
we obtain, from Eq.~(\ref{eq:Vk3}):
\begin{equation}
\langle V_N \rangle =2\,\sum_{k=1}^{N} e_k. \label{eq:Vexcprob}
\end{equation}
Eq.~\ref{eq:Vexcprob} embodies the core of the reasoning followed here and hints at how one can generalize it to continuous-time processes.
\medskip

Now, the probability for a random walk with $k$ steps to lie on one side only of the line joining its end points can be computed from Baxter's combinatorial lemma~\cite{Bax}, or by considering the following question: given a planar random walk with $k$ steps, calling $\mathcal{L}$ the line joining the endpoints of the walk, what is the joint probability distribution of the number~$T^{(1)}$ of steps lying on one side of $\mathcal{L}$ and the number~$T^{(2)}$ of steps lying on the other side?\\
The walk's jumps being independent and identically distributed, all pairs $(T^{(1)},T^{(2)})$ (satisfying $1\leq T^{(1)}, T^{(2)}\leq k-1$ and $T^{(1)}+T^{(2)}=k-1$) are equiprobable. Only two of these pairs correspond to the walk perfoming an excursion (that is never crossing the line joining its endpoints). These are $(k-1,0)$ and $(0,k-1)$ and therefore the sum of the probabilities associated with an excursion event is
\begin{equation}
2\,e_k=\frac{2}{k}.\label{eq:DTExc}
\end{equation}

Substituting this value for $e_k$ in Eq.~(\ref{eq:Vexcprob}), one obtains:
\begin{equation}
\langle V_N \rangle = 2 \sum_{k=1}^{N} \frac{1}{k}, \label{eq:VDT}
\end{equation}
which is the same result as in \cite{Bax}.

\section{Continuous-time case}
\label{CT}
Let us now consider a planar Lévy process, that is a Markov process $Z(t)=(x(t),y(t))$ in the plane ($t \in [0,T]$ for some fixed $T>0$), where $x$ and $y$ are independent, continuous-time processes with independent, stationary increments. We write $\Gamma ^{(0)} _t$ for the process's path from time $0$ up to time $t$ and $\partial \mathcal{C}_t$ for the boundary of its convex hull. Without loss of generality, we assume $Z(0)=(0,0)$.\\ For any given $s \in [0,T]$ and any $\tau \in [0,T-s]$ we want to compute the probability that the line segment joining $Z(\tau)$ and $Z(\tau + s)$ will appear on~$\partial \mathcal{C}_T$.

Of course, unlike in the discrete-time case, the polygonal nature of $\partial \mathcal{C}_T$ is not obvious. However, in the Brownian case, it is known that $\partial \mathcal{C}_T$ consists almost surely of countably many straight line segments~\cite{Ev,CHM}. This may not hold in the general case of Lévy processes, but, notwithstanding the existence of other types of subsets in $\partial \mathcal{C}_T$, one can endeavour to count edges of the type described above. This is what we do here, as a first step toward describing the structure of~$\partial \mathcal{C}_T$ for non-Brownian processes in the plane (it seems that ours is the first result in this direction).

\begin{figure}
\begin{center}
\includegraphics[scale=0.3]{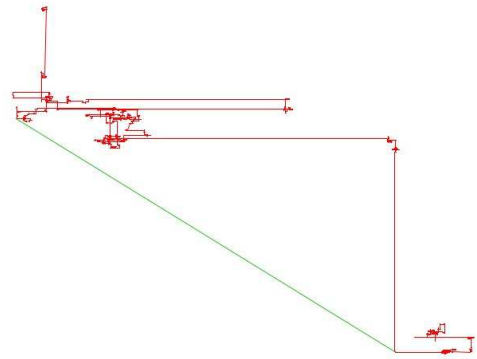}
\end{center}
\caption{\label{fig:CauSedge}Example of an $s$-edge: a line segment (bottom left) appearing on the convex hull of a Cauchy process's sample path, joining two points of the path visited at times $\tau$ and $\tau +s$.}
\end{figure}

\subsection{Number of $s$-edges}
Similarly to what was done in the discrete-time case, we use the Markov property of the path to express $\langle V^{(s)}\rangle\, ds$, the average number of $s$-edges joining two points separated by a time-lapse belonging to~$[s,s+ds]$, as a product of three terms corresponding to three independent parts of the path: before some $\tau \in [0,T-s]$, between $\tau$ and $\tau +s$, after $\tau +s$~(see~Figure~\ref{fig:CauSedge}) This yields:
\begin{equation}
\langle V^{(s)} \rangle\, ds = 2\,\int _{\tau=0}^{T-s} p^{0}_{(\tau)}\, f(\tau,s)\, p^{(\tau +s)}_{T}\, d\tau\,ds, \label{eq:Vs}
\end{equation}
where the three factors in the integrand are associated respectively with the probabilities of the following independent events: (a) $\Gamma ^{(0)} _\tau$ lies on one side only (say side~$1$) of a line~$\mathcal{L}$ given that the path hits~$\mathcal{L}$ at time $\tau$; (b) $\Gamma ^{(\tau)} _{\tau +s}$ is an excursion away from~$\mathcal{L}$ (in side~$1$) and is pinned on the line at times $\tau$ and $\tau+s$; (c) $\Gamma ^{(\tau +s)} _T$ lies on side~$1$ of~$\mathcal{L}$ given that it is pinned on the line at time~$\tau +s$. 

Since the process has stationary, independent increments, the excursion probability appearing as the middle factor will, just as in the discrete-time case, not depend on $\tau$:
\begin{equation}
\forall \tau \in [0,T-s],\quad f(\tau,s)=f(s),\nonumber
\end{equation}
so that Eq.~(\ref{eq:Vs}) becomes:
\begin{equation}
\langle V^{(s)} \rangle\, ds = 2\,\left(\int _{\tau=0}^{T-s} p^{0}_{(\tau)}\ p^{(\tau +s)}_{T}\, d\tau\right)\, f(s)\, ds.\label{eq:Vs2}
\end{equation}
Given the Markovian and time-homogeneous nature of the process, the integrand can again be interpreted as giving the distribution of the time $\tau$ at which a similar process of duration $T-s$ hits a line given that it does hit that line but never crosses it. The integral in Eq.~(\ref{eq:Vs2}) will therefore be equal to $1$, which leads to the following formula for the average number of edges joining points separated by time-lapses in $[\Delta \tau _1,\Delta \tau _2]$:
\begin{eqnarray}
\langle V(\Delta \tau _1,\Delta \tau _2)\rangle &=& \int_{\Delta \tau_1}^{\Delta \tau_2}\,\langle V^{(s)} \rangle\,d s\nonumber\\
&=& 2\,\int_{\Delta \tau_1}^{\Delta \tau _2}\,f(s)\,d s.\label{eq:VCTexcprob}
\end{eqnarray}

\subsection{Continuous edge-excursion formula}
Now, just as in the discrete-time case, $f(s)$ will correspond to the probability that a 2-dimensional process of duration~$s$ performs an excursion from the line joining its two endpoints: i.e. drawing the line through the initial and final points of a path of duration~$s$, $f(s)$ corresponds to the probability that the sojourn time of the process on a given side of the line is equal to the full duration~$s$. Writing $\rho_s(u)$ for the density of the sojourn time of a process of duration~$s$ on one side of the line through its endpoints (note that $\rho_s(u)=0$ if $u \notin [0,s]$) and substituting in Eq.~(\ref{eq:VCTexcprob}), we obtain our main result:
\begin{equation}
\langle V(\Delta \tau _1,\Delta \tau _2)\rangle =2\,\int_{\Delta \tau_1}^{\Delta \tau _2}\,\rho_s(s)\,ds. \label{eq:VCTsoj}
\end{equation}
This is the general, continuous-time equivalent to Eq.~(\ref{eq:Vk3}).

\subsection{Uniform sojourn times}
From~Eq.~(\ref{eq:VCTsoj}), we need to compute $\rho _s$, that is the sojourn time density on a given side of a line for a planar Lévy process with duration $s$, constrained to start from the line at time~$0$ and to hit the same line again at time~$s$. Moving into a coordinate system where that line coincides with the $x$-axis, and the starting point of the process with the origin, we want to know the density for the sojourn time, say, in~$\mathbb{R}^{+}$ of the $y$ coordinate. As a linear combination of independent one-dimensional Lévy processes, this coordinate will be performing a one-dimensional Lévy process (in general not independent of the $x$-process, but that does not matter here), constrained to come back to $0$ at the end of its time interval --- this is called a Lévy bridge.

A well-known result by Lévy \cite{Lev2} is that for linear Brownian motion, the time spent in $\mathbb{R}^{+}$ or $\mathbb{R}^{-}$ follows the same distribution (generally called the Arcsine law) as the time at which the motion attains its maximum (or equivalently its minimum). A similarly well-known fact is that these two distributions are also equal in the case of a Brownian bridge, and thus the sojourn time distribution for a Brownian bridge is uniform over its full duration~$s$, namely it is a constant equal to $\frac{1}{s}$ for sojourn times $t^{+}\in [0,s]$ and equal to $0$ otherwise. This result is linked to a class of combinatorial theorems in probability theory (the so-called Ballot theorems~\cite{ABR}) and, remarkably, it extends to Lévy bridges and other types of processes with cyclically exchangeable increments~\cite{Kn,Kal,Yor}. More precisely, the sojourn time distribution of a Lévy bridge will be uniform if the Fourier transform of the underlying free Lévy process is integrable~(see~\cite{Kn} for details). The condition holds in particular for symmetric $\alpha$-stable processes, which include Brownian motion ($\alpha=2$), the Cauchy-Lorentz process ($\alpha=1$), and other processes generically called Lévy flights ($0<\alpha<2$), as well as for certain compound Poisson processes with drift. Therefore, in all those cases at least, the uniform law for sojourn times combined with~Eq.~(\ref{eq:VCTsoj}) will lead to:
\begin{eqnarray}
\langle V(\Delta \tau _1,\Delta \tau _2)\rangle &=& 2\,\int_{\Delta \tau_1}^{\Delta \tau _2}\,\frac{1}{s}\,ds \nonumber \\
&=& 2\log \left( \frac{\Delta \tau_2}{\Delta \tau_1} \right). \label{eq:VCTlog}
\end{eqnarray}

\section{Numerical simulations}
\label{NS}

\begin{figure}
\begin{center}
\includegraphics[scale=0.3]{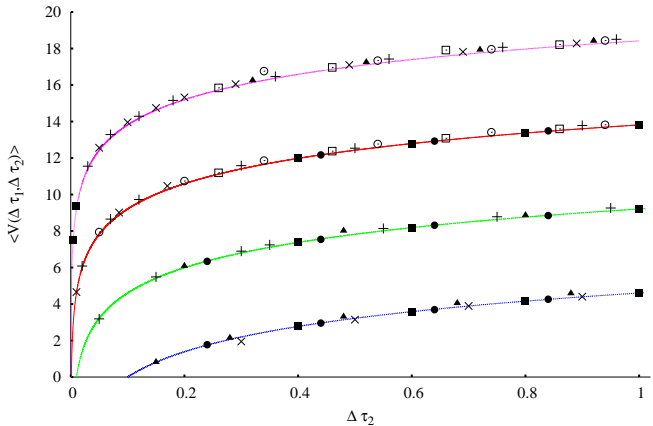}
\end{center}
\caption{\label{fig:NumSim}Computer simulation results (points) compared with the exact formula from Eq.~(\ref{eq:VCTlog}) (solid lines) giving the average number of $s$-edges with $s\in[\Delta \tau _1,\Delta \tau _2]$, for a range of values of~$\Delta \tau_2$ and $\Delta \tau_1$ ($\Delta \tau_1 = 10^{-1}$ on the bottom curve, $10^{-2}$ on the second  curve, $10^{-3}$ on the third one and $10^{-4}$ on the top curve). Symbols correspond to computer simulations with $10^{4}$ paths for: Brownian motion with diffusion constant set to $1/2$ (filled squares) or $50$ (filled circles), Cauchy-Lorentz process (filled triangles), symmetric $\alpha$-stable processes with $\alpha=0.75$ (squares) and $\alpha=1.5$ (circles), compound Poisson processes with drift and with normal (pluses) or Cauchy (crosses) jump variables.}
\end{figure}

To illustrate and confirm the validity of Eq.~(\ref{eq:VCTlog}), we performed numerical simulations for four types of planar Lévy processes: (i) Brownian motion, (ii) Cauchy-Lorentz process (with increments' distribution~$t/\pi(x^2+t^2)$), (iii) two other symmetric $\alpha$-stable processes with $\alpha=3/4$ and $\alpha=3/2$, and (iv) compound Poisson processes with drift (given by $Z(t)=\mu t+\sum_{i=0}^{N(t)}D_i$, where $\mu$ is the drift, $N(t)$ is Poisson distributed with parameter $\lambda t$ for some $\lambda >0$ and the $D_i$'s are independent, identically-distributed variables).

We ran, in each case, $10^{4}$~sample paths' realisations and computed the average number of $s$-edges on the convex hull of the paths, with $s\in[\Delta \tau _1,\Delta \tau _2]$ for a range of values of~$\Delta \tau_1$ and $\Delta \tau_2$. As can be seen on Figure~\ref{fig:NumSim}, the agreement with the exact values given by $2\log \left( \frac{\Delta \tau_2}{\Delta \tau_1} \right)$ is very good, thereby illustrating the validity of~Eq.~(\ref{eq:VCTlog}).

\section*{Conclusion}
Eqs.~(\ref{eq:VCTsoj}) and (\ref{eq:VCTlog}) are, to the best of our knowledge, the first general, analytical formulae related to the average number of edges appearing on the convex hull of generic continuous-time planar processes. They lead to the existence of universality concerning this average number, as well as a certain time-scale invariance. More work is needed to explore the precise extent of the universality class identified here, but we have seen that it contains at least symmetric $\alpha$-stable processes and certain compound Poisson processes with drift.

The derivation presented here relies strongly on the Markov nature of Lévy processes, and some intrinsic symmetry associated with one-dimensional Lévy bridges. Applications, in particular those to polymer physics, will require further developments, especially toward non-Markov processes (incl. non-Markovian anomalous diffusion processes~\cite{LGE}, random-acceleration processes~~\cite{RMR} and constrained processes like excluded-volume or non self-intersecting processes~\cite{DoiEdw,dGe}): will there be other universality classes or just case-by-case formulae for the average number of edges appearing on the convex hull of sample paths? Also relevant will be the study of the global convex hull of multiple dependent or independent paths~\cite{BMCHEdges,DMRZ}.

Note that, should any of these situations lead to a non-uniform sojourn-time distribution for the corresponding one-dimensional bridge-type configurations, then the final formula could be drastically different, \textit{eg} power-law instead of logarithmic, pointing to a much faster growth of the number of edges on the boundary of the convex hull.

\bibliography{SPCHEdges2.bib}

\end{document}